\documentclass[%
reprint,
superscriptaddress,
amsmath,amssymb,
aps,longbibliography,
prx
]{revtex4-2}

\usepackage{graphicx}%
\usepackage{verbatim}
\usepackage[version=4]{mhchem}
\usepackage{xcolor}
\usepackage{physics}
\usepackage[normalem]{ulem}
\usepackage{pdfpages}
\usepackage{pgffor}
\usepackage{booktabs}
\usepackage{hyperref}

\usepackage[capitalise]{cleveref}
\usepackage{bm}%

\makeatletter
\AtBeginDocument{\let\LS@rot\@undefined}
\makeatother

\newcommand{\LiPS}{\ce{Li3PS4}}

\usepackage[normalem]{ulem}
\definecolor{tangerine}{rgb}{0.944,0.522,0}
\definecolor{verde}{rgb}{0.,0.6,0}
\definecolor{rosso}{rgb}{0.9,0.0,0.2}
\definecolor{magenta}{rgb}{0.9,0.2,0.9}

\newif\ifhighlight

\newcommand{\highlight}{\highlighttrue}
\highlight
\newcommand{\editor}[2]{%
  \expandafter\newcommand\csname #1note\endcsname[1]{%
    \textcolor{#2}{(\textbf{#1note:} \textsc{##1})}}%
  \expandafter\newcommand\csname #1\endcsname[1]{%
    \ifhighlight\textcolor{#2}{##1} \else ##1\fi}%
  \expandafter\newcommand\csname #1cancel\endcsname[1]{%
    \ifhighlight\textcolor{#2}{\sout{##1}}\fi}%
  \expandafter\newcommand\csname #1change\endcsname[2]{%
    \ifhighlight\textcolor{#2}{\sout{##1} ##2}\else ##2\fi}%
  \newenvironment{#1text}{\ifhighlight\color{#2}\fi}{\color{black}}
}

\editor{FG}{red}
\editor{resub}{blue}
\editor{DT}{verde}
\editor{LG}{orange}

\begin{document}

\title{Thermal conductivity of \LiPS~solid electrolytes with \textit{ab initio} accuracy}

\author{Davide Tisi}
\affiliation{Laboratory of Computational Science and Modeling, Institut des Mat\'eriaux, \'Ecole Polytechnique F\'ed\'erale de Lausanne, 1015 Lausanne, Switzerland}

\author{Federico Grasselli}
\affiliation{Laboratory of Computational Science and Modeling, Institut des Mat\'eriaux, \'Ecole Polytechnique F\'ed\'erale de Lausanne, 1015 Lausanne, Switzerland}

\author{Lorenzo Gigli}
\affiliation{Laboratory of Computational Science and Modeling, Institut des Mat\'eriaux, \'Ecole Polytechnique F\'ed\'erale de Lausanne, 1015 Lausanne, Switzerland}

\author{Michele Ceriotti}
\email{michele.ceriotti@epfl.ch}
\affiliation{Laboratory of Computational Science and Modeling, Institut des Mat\'eriaux, \'Ecole Polytechnique F\'ed\'erale de Lausanne, 1015 Lausanne, Switzerland}

\date{\today}%

\begin{abstract}
{The vast amount of computational studies on electrical conduction in solid-state electrolytes is not mirrored by comparable efforts addressing thermal conduction, which has been scarcely investigated despite its relevance to thermal management and (over)heating of batteries. 
The reason for this lies in the complexity of the calculations: 
on one hand, the diffusion of ionic charge carriers makes lattice methods formally unsuitable, due to the lack of equilibrium atomic positions needed for normal-mode expansion. On the other hand, the prohibitive cost of large-scale molecular dynamics (MD) simulations of heat transport in large systems at \textit{ab initio} levels has hindered the use of MD-based methods. In this paper, we leverage recently developed machine-learning potentials targeting different \textit{ab initio} functionals (PBEsol, r$^2$SCAN, PBE0) and a state-of-the-art formulation of the Green-Kubo theory of heat transport in multicomponent systems to compute the thermal conductivity of a promising solid-state electrolyte, \LiPS,~in all its polymorphs ($\alpha$, $\beta$, and $\gamma$). By comparing MD estimates with lattice methods on the low-temperature, nondiffusive $\gamma$-\LiPS, we highlight strong anharmonicities and negligible nuclear quantum effects, hence further justifying MD-based methods even for nondiffusive phases. Finally, for the ion-conducting $\alpha$ and $\beta$ phases, where the multicomponent Green-Kubo MD approach is mandatory, our simulations indicate a weak temperature dependence of the thermal conductivity, a glass-like behavior due to the effective local disorder characterizing these Li-diffusing phases.}

\end{abstract}

\maketitle

\section{Introduction}

Thermal management is recognized as a major issue in both the design and practical manufacturing of modern batteries: insufficient or inhomogeneous thermal dissipation leads to overheating, particularly during rapid charging processes. Such overheating can, in turn, trigger severe events, like partial melting or explosions, or, more generally, degradation, non uniform charge conduction, and dendrite formation \cite{Rao2022,Thakur2023,Wang2023}. Given these circumstances, it is entirely justified to consider thermal runaway as ``the key scientific problem in battery safety research'' (verbatim from Ref.~\onlinecite{Feng2018}). 
Despite the need to study, in a quantitative way, the concurring mechanisms of heat dissipation (via ionic Joule effect) and charge flow at operating regimes, \textit{thermal} transport in solid-state electrolytes (SSEs), which constitute the core of modern Li-ion batteries, has been only scarcely investigated, especially when compared to \textit{charge} transport \cite{deshpande2020roadmap,KUDU2022168,HOMMA201153,kimura2023,de2018analysis,forrester2022,staacke2021role,Staacke2022,guo_artificial_2022,harper2024tracking,miyagawa2024accurate,Ryo2024}.
The interest in performing measurements of the thermal conductivity for several classes of SSEs has grown only very recently \cite{Cheng2021,wu2022,Boger2023}. Computational studies, which are abundant for other classes of ionic materials \cite{Ohtori2009,KIM2015247,Zhou2018,Knoop2023}, are even fewer than experimental ones and often rely on very crude models of microscopic thermal transport, which completely neglect the effects of vacancies, defects and, most importantly, ionic diffusion. This is due to the complexity of the calculations of thermal conductivity in SSEs: On the one hand, the absence of well-defined equilibrium atomic positions makes lattice methods---even when adapted to deal with disorder and defects \cite{Isaeva2019,Simoncelli2019,Caldarelli2022,Fiorentino2023}---formally not suitable for SSEs, characterized by Li-ion diffusion occurring at the same time scale of heat transfer.
On the other hand, 
{although considerable theoretical efforts in the community have recently made \textit{ab initio} simulations of heat transport from equilibrium molecular dynamics (MD) possible }\cite{marcolongo2016microscopic,marcolongo2021qeheat,KangWang2017,Carbogno2017}, the computational cost makes them impractical for large-scale simulations. In the last decade, the rise of machine-learning potential (MLPs) in combination with the Green-Kubo (GK) theory of linear response \cite{Green,Kubo} allowed the community to study the thermal properties of several complex and large systems \cite{Sosso2012,Korotaev2019,QIAN2019100140,Li2020,Shimamura2020,LI2020100181,Mangold2020,Liu_2021,fan2021,Verdi2021,Langer2023,Xu2023,Sun2023,Pegolo2022,Pegolofrontiers2024,dong2024molecular,Tisi2021}.\\
In this paper, we address thermal conduction in a paradigmatic SSE, lithium ortho-thiophosphate (\LiPS), a promising candidate for next-generation batteries thanks to the large ionic conductivity and wide electrochemical stability of its superionic polymorphs (the $\alpha$ and $\beta$ phases) \cite{KUDU2022168,HOMMA201153,kimura2023,de2018analysis,forrester2022,staacke2021role,Staacke2022,guo_artificial_2022}. 
We leverage machine-learning potentials~\cite{Gigli2023} and {the extension of the GK theory of linear response} to multicomponent systems \cite{bertossa2019} to compute the thermal conductivity $\kappa$ of the different \LiPS\, polymorphs {(i.e., the Li-ion diffusive $\alpha$ and $\beta$ phases, as well as the nondiffusive $\gamma$ phase)} over a wide range of temperatures. The use of MLPs also allows us to compare estimates of $\kappa$ from different target levels of density functional theory (DFT), namely the generalized-gradient-approximation (GGA) PBEsol~\cite{perdew1996generalized,PerdewPBEsol}, the meta-GGA regularized SCAN (r$^2$SCAN)~\cite{r2SCAN} and the hybrid PBE0~\cite{adamo1999toward} functionals. 

\section{Theory and Methods}
\subsection{Machine learning models}

We use MLP models from our recent study on charge transport in \LiPS~\cite{Gigli2023}: MLPs are constructed from kernel ridge regression, using the power spectrum of the smooth overlap of atomic positions (SOAP) \cite{bart+13prb} as features. The MLP model is built to fit the total potential energy $E$, and force components $\{\bm{F}_{i=1}^N\}$, where $N$ is the number of atoms, over a training set of representative structures. Target properties are expressed as sums of local, atom-centered contributions predicted through kernel ridge regression,
\begin{equation}
\begin{split}
    E = \sum_{i \in A} \epsilon_i(A_i) &= \sum_{i \in A} \sum_{j \in M} w_j \, \mathrm{k}(A_i, M_j) \\
    \bm{F}_i &= -\mathbf{\grad}_i E \label{eq:forcesridge}
\end{split}
\end{equation}
Here, $A_i$ represents the environment surrounding atom $i$ in structure $A$, $\epsilon_i(A_i)$ the potential energy associated with it, $w_j$ are the weights to be optimized, and the kernel $\mathrm{k}(A_i, M_j)$ measures the correlation between the environment $A_i$ and an environment $M_j$, taken from a representative set, the so-called sparse set $M$.
{We have used a quadratic kernel, $\mathrm{k}(A_i, A_j) = |\mathbf{p}_i \cdot \mathbf{p}_j|^2$, where $\mathbf{p}_i$ is the SOAP feature of the local environment $A_i$. Further details and specific hyperparameters of these Gaussian approximation potentials \cite{Bartok2010} can be found in the Supporting Information of Ref.~\cite{Gigli2023}.}
Post-GGA models are trained using a two-level scheme \cite{Lilienfeld2015,Zaspel2019} whereby the difference with respect to a PBEsol ML baseline is learned on a representative subset of structures on which post-GGA calculations are performed. This learning strategy avoids running thousands of expensive post-GGA \textit{ab initio} calculations while retaining good accuracy. Further details on data set generation and training details can be found in  Ref.~\onlinecite{Gigli2023}.
Since thermal conduction is intertwined with the lattice elastic properties of a material, we have tested the reliability of our MLP {in reproducing the bulk modulus of \LiPS~via Murnaghan's equation of state (MEoS) \cite{Murnaghan1937,Murnaghan1944,Birch1947}} finding an agreement {within $3~$GPa, which is less than 10\% of the typical values for this relatively soft material}. See Supplemental Material (SM)~\cite{suppmat}, Sec.~II, for details. In Sec.~IV of the SM we discuss the phonon bands and vibrational density of states obtained with our MLP, finding good agreement with those computed \textit{ab initio} within density functional perturbation theory \cite{BaroniRev2001}.

\subsection{Multicomponent Green-Kubo theory}

Heat transport is quantified, in the linear response regime, by the thermal conductivity, $\kappa$, defined as the ratio between the heat flux and the negative temperature gradient in the absence of any convection. 
The GK theory of linear response \cite{Green,Kubo} provides a rigorous framework to compute $\kappa$ from equilibrium MD (EMD) simulations in terms of the stationary time series of properly defined microscopic fluxes sampled along the EMD trajectory. 
For one-component systems, solids and molecular fluids characterized by a single, nondissociated molecular type, the prescription of vanishing convection is trivially satisfied in MD simulations performed in the barycentric reference frame, since the center of mass of each atomic species does not diffuse. In particular, for an isotropic system of $N$ interacting particles, the one-component GK expression reads:
\begin{equation}\label{eq:GKeq}
    \kappa=\frac{\Omega}{3k_\mathrm{B} T^2}\int_0^{\infty} \langle \bm{J}_e(\Gamma_t) \cdot \bm{J}_e(\Gamma_0) \rangle \, \mathrm{d}t,
\end{equation}
where $k_\mathrm{B}$ is the Boltzmann constant, $T$ the temperature and $\Gamma_t$ indicates the evolution at time $t$ of a point in phase space from the initial condition $\Gamma_0$, over which the equilibrium average $\langle \cdot \rangle$ is performed. The energy flux $\bm{J}_e$ is defined as \cite{Irving1950}:
\begin{align}
    \bm{J}_e &= \frac{1}{\Omega}\sum_{i=1}^{N} \left[ e_i \bm {v}_i -\sum_{j=1}^{N} (\bm r_i - \bm r_j) \frac{\partial \epsilon_j}{\partial \bm{r}_i}\cdot \bm {v}_i \right] \label{eq:jenergy}
\end{align}
where $\bm{r}_i$, $\bm{v}_i$ are the atomic positions and velocities, respectively; $e_i=\frac{1}{2}m_iv_i^2+\epsilon_i$ is the atomic energy, sum of a kinetic contribution ($m_i$ is the atomic mass) and a local potential energy term, $\epsilon_i$. In the MLP picture, $\epsilon_i$ is given by the atomic partition of the potential energy of the ML model, as in \cref{eq:forcesridge} \cite{Tisi2021}. 
Recently, Langer \textit{et al.}~\cite{Langer2023,Langer2023stress} have also extended the definition of ML energy flux to semilocal, message-passing MLPs, and implemented it via automatic differentiation techniques.

\begin{figure}[t!]
    \centering
    \includegraphics[width=\columnwidth]{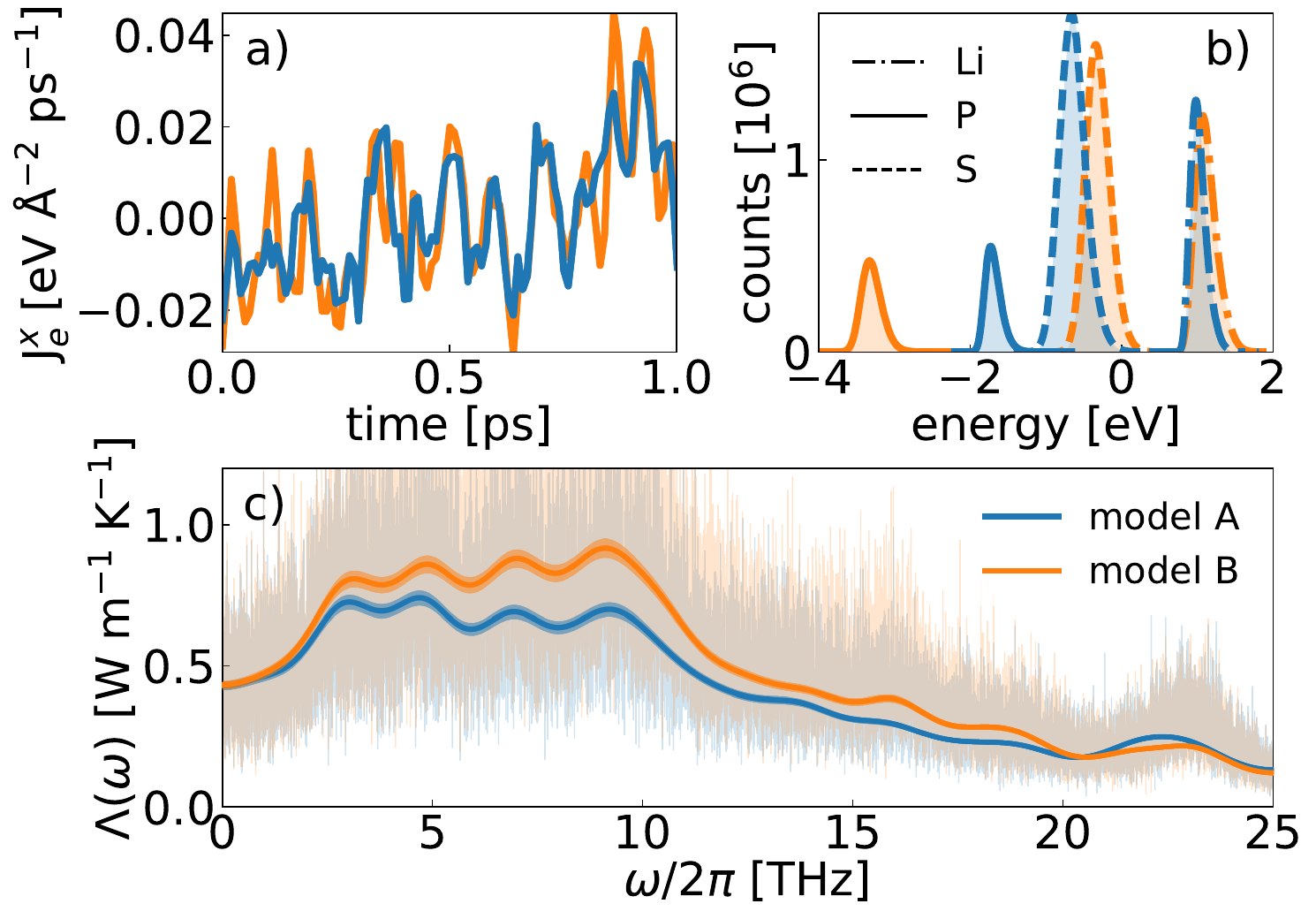}
    \caption{Nonuniqueness of energy flux and invariance principles of thermal conductivity. (a) $x$ Cartesian component of the energy flux $\bm{J}_e$ predicted, along a same segment of trajectory, by two models, here labeled A (blue) and B (orange), of a committee of PBEsol MLPs. (b) Distribution of atomic energies per species for the same two models over the same trajectory. Both shape and position differ between models A and B. (c) Reduced power spectra (pale, noisy background lines) and their cepstral filter (thick solid line with uncertainty) for models A and B. The values at $\omega\to 0$, representing the estimate of $\kappa$ for models A and B, coincide. All results are obtained by recomputing the local energies and heat fluxes with models A and B on a trajectory generated by model A for the $\alpha$ phase equilibrated at $T=650$~K and zero pressure.}
    \label{fig:committee_and_invariance}
\end{figure}

\begin{figure}[h]
    \centering
    \includegraphics[width=\columnwidth]{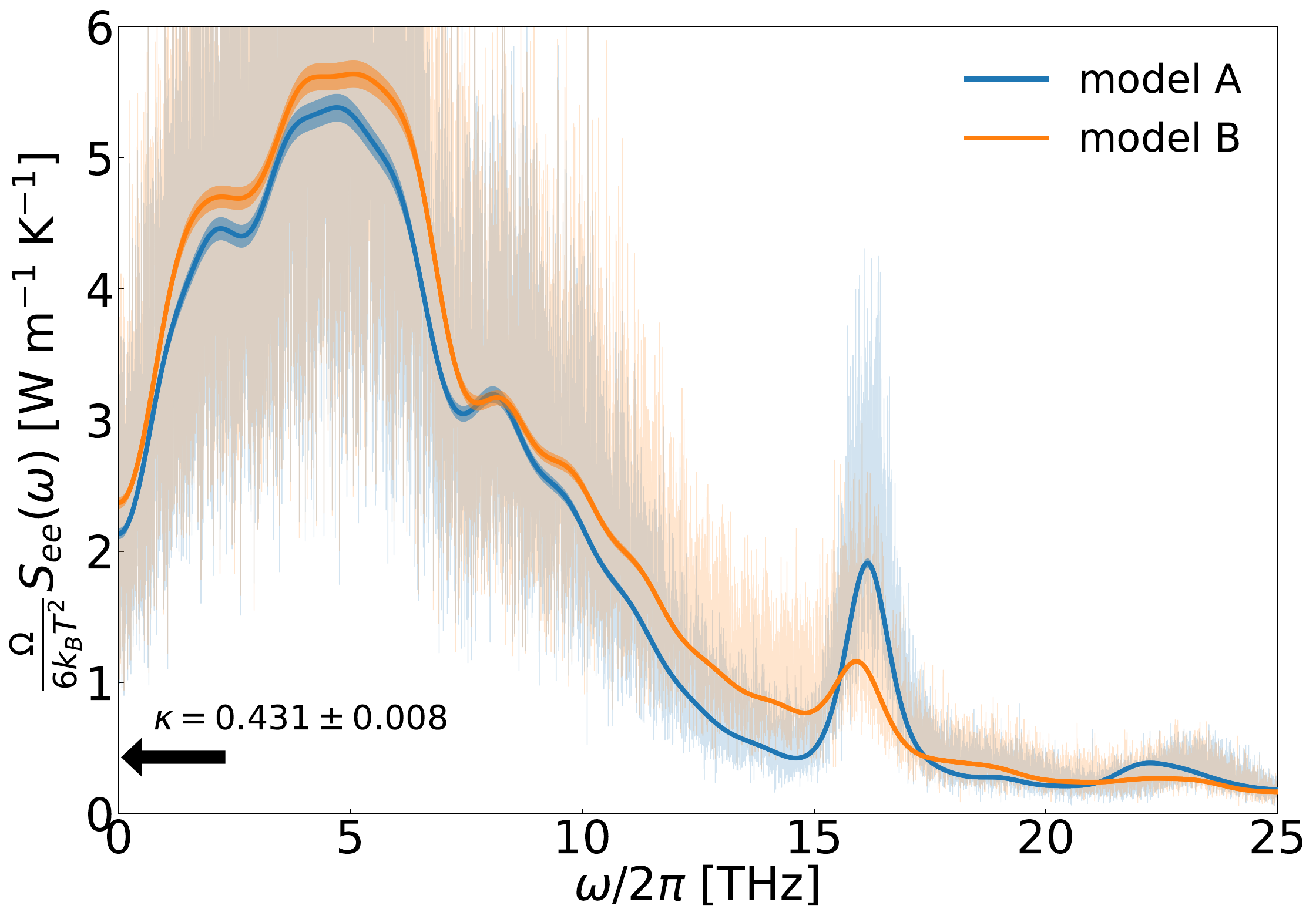}
    \caption{Effect of incorrect use of $S_{ee}(\omega)$ instead of the full $\Lambda (\omega)$. $S_{ee}(0)$ not only depends on the ML model, but it is also very different from the correct value, taken from \cref{fig:committee_and_invariance} and represented by the black arrow. }
    \label{fig:single_component}
\end{figure}

Nevertheless, for a multicomponent system such as the conductive phases of \LiPS, Eq.~\eqref{eq:GKeq} cannot be applied as is: in fact, in a standard EMD simulation, keeping the center of mass of the entire system fixed does not imply that the centers of mass of each atomic species do not diffuse, and the prescription of vanishing convection is not satisfied \cite{Lindan1991}. Therefore, a theoretically justified and computationally robust method must be used to remove the coupling between the energy flux, $\bm{J}_e$ and the mass fluxes $\bm{J}_X = \tfrac{1}{\Omega} m_X \sum_{i\in X} \bm{v}_i$ of atomic species $X=\text{Li, P, or S}$. 
{A multivariate technique to do so was developed in Ref.~\onlinecite{bertossa2019}}:
$\kappa$ is estimated in terms of the zero-frequency limit of the so-called \textit{reduced} power spectral density (PSD), i.e., the PSD $S_{ee}(\omega)$ of the energy flux that is obtained after removing its coupling to the
mass fluxes: 
\begin{equation}\label{eq:S_omega}
\begin{split}
    \kappa &= \lim_{\omega\to 0} \Lambda(\omega) \\
    \Lambda(\omega) &\equiv \frac{\Omega}{6 k_\mathrm{B} T^2} \underbrace{[S_{ee}(\omega) - S_\mathrm{coupl.}(\omega)]}_{\text{reduced PSD}}
\end{split}
\end{equation}
Here,
\begin{equation}
\begin{split}
    S_{\mathrm{coupl.}}&\equiv
        \begin{pmatrix}
        S_{\text{Li} e} & S_{\text{P} e}
        \end{pmatrix}
        \begin{pmatrix}
        S_{\text{LiLi}} & S_{\text{LiP}} \\
        S_{\text{PLi}} & S_{\text{PP}}
        \end{pmatrix}^{-1}
        \begin{pmatrix}
        S_{e \text{Li}} \\
        S_{e \text{P}}
        \end{pmatrix}
\end{split}
\end{equation}
and $S_{ab}(\omega) = \int_{-\infty}^{\infty} \langle \bm{J}_a(\Gamma_t) \cdot \bm{J}_b(\Gamma_0) \rangle e^{-i\omega t} \, \mathrm{d}t$
is the cross-PSD between fluxes $\bm{J}_a$ and $\bm{J}_b$, with $a,b\in \{e,\mathrm{Li},\mathrm{P}\}$ \footnote{The mass flux of sulfur atoms does not appear explicitly, as it can be expressed in terms of $\bm{J}_\mathrm{Li}$ and $\bm{J}_\mathrm{P}$, because of total momentum conservation}. Notice that for solids and single-component systems, $\lim_{\omega\to 0} S_{\mathrm{coupl.}} = 0 $, and Eq.~\eqref{eq:S_omega} reduces to Eq.~\eqref{eq:GKeq}, thanks to the Wiener-Khintchine theorem \cite{Wiener1930,Khintchine1934}. See Refs.~\onlinecite{baroni2020heat,Grasselli2021} for further details.
An important remark is necessary: the partitioning of the total potential energy $E(A)$ into local contributions $\epsilon_i$ is not unique. In fact, two models that differ by a small change in the hyperparameters and/or training set, will lead to equivalent total energies and forces, but different $\epsilon_i$. {Nevertheless, GK theory still allows to uniquely define $\kappa$ thanks to recently discovered \textit{invariance principles} of transport coefficients, according to which $\kappa$ is largely independent of the definition of microscopic energies \cite{Marcolongo2020,Ercole2016,Grasselli2021,Tisi2021}}
This can be verified by explicit construction of a \textit{committee} of MLPs via dataset subsampling \cite{musil_fast_2019}, %
through which we can quantify how much local predictions are dependent on the model's details~\cite{chong2023robustness}.
Such a model dependence of $\epsilon_i$ implies that different MLPs inevitably lead to different instantaneous values of the energy flux, $\bm{J}_e$, which is defined from local quantities. This is explicitly shown in \cref{fig:committee_and_invariance} for a simulation of the $\alpha$ phase of \LiPS~at $T=650$ K, which is superionic and characterized by significant Li-ion diffusion. We consider a single EMD trajectory driven by a model (here labeled A) of a committee of MLPs obtained from subsampling the whole PBEsol dataset. We then compute the energy flux via Eq.~\eqref{eq:jenergy}, over this trajectory, for model A, as well as for other members of the same committee. We use the same trajectory to single out the role of model-dependent estimates of $\bm{J}_e$ from the effects due to sampling independent EMD trajectories, each run with a different member of the MLP committee. Despite the slight procedural abuse \footnote{Equation~\eqref{eq:jenergy} is obtained by enforcing Newton's second law, hence the use of a single trajectory is justified only when the spread in the forces predicted by the committee members vanishes.}, this approach allows us to test the invariance principles of $\kappa$ more directly. 
Figure~\ref{fig:committee_and_invariance} shows that different members of the committee give, on the same segment of trajectory, different estimates for $\bm{J}_e$ (panel a), as well as for the distribution of atomic energies per species (panel b). Nonetheless, the $\omega\to 0$ limit of the reduced PSD, here estimated via multivariate cepstral analysis~\cite{Ercole2017,bertossa2019}, is independent of the specific MLP committee member, as clearly shown in Fig.~\ref{fig:committee_and_invariance}(c). An analogous analysis on more members of the committee gives the same result, see Fig.~S5 within the SM~\cite{suppmat}.
{It is worth remarking that whenever MLP models are trained for multicomponent systems with a fixed dataset stoichiometry, as in the case of \LiPS, the average per-atom potential energies (and not only the shape of their distributions) is model dependent, see Fig.~\ref{fig:committee_and_invariance}(b). This fact and the presence of diffusing species induce \textit{model-dependent}, convective-like contributions to the heat flux that are responsible for different and incompatible values of $S_{ee}(0)$, as shown in the $\omega\to 0$ region of Fig.~\ref{fig:single_component}. Therefore, these values cannot be taken as meaningful estimates of $\kappa$, and are indeed several times larger than the correct value (black arrow), obtained from the multivariate formulation. Only the latter is able to remove the coupling of these convective contributions from heat conduction via the calculation of the \textit{reduced} PSD.}

The cepstral analysis of time series \cite{Ercole2017}, employed to filter the noisy reduced PSD of Fig.~\ref{fig:committee_and_invariance}(c) {and \cref{fig:single_component}}, is a powerful filtering technique that has been already applied to estimate the thermal conductivity of complex materials \cite{grasselli2020heat,Pegolo2022,Pan2024}, as well as other transport coefficients \cite{malosso2022viscosity,PegoloMS2023}. We employ its implementation in the \textsc{SporTran} code \cite{Ercole2022sportran}. 
The calculation of $\bm{J}_e$, as shown in \cref{eq:jenergy}, consists of two terms: the first term depends only on $e_i$ and the velocities and is trivial to compute; the second term depends on the so-called \textit{atomic virial}, defined by %
\begin{equation}
    \mathrm{W}_{i,\mu\nu} \equiv - \sum_j ( r_{i,\mu} -  r_{j,\mu} ) \frac{\partial \epsilon_j}{\partial r_{i,\nu}}
\end{equation}
with $\mu, \nu \in \{x,y,z\}$. This is implemented in \textsc{librascal} \cite{librascal} by direct application of the chain rule to the SOAP features. Notice that, contrary to the total virial, the atomic virial need \textit{not} be symmetric and all the nine Cartesian components must be stored to avoid miscalculations of $\kappa$ (see Refs.~\cite{Fan2015,Surblys2019}). {In particular, from the implementation standpoint, only pair potentials $v(|\bm{r}|)$ prescribe that $\mathrm{W}_{i,\mu\nu} = \mathrm{W}_{i,\nu\mu}$. In fact, 
\begin{equation}
    \frac{\partial v(|\bm{r}_{ij}|)}{\partial r_{i,\nu}} = \frac{\partial v(|\bm{r}_{ij}|)}{\partial r_{ji,\nu}} = \frac{1}{|\bm{r}_{ji}|} \frac{\partial v(|\bm{r}_{ij}|)}{\partial |\bm{r}_{ji}|} r_{ji,\nu}, 
\end{equation}
where $\bm{r}_{ji} \equiv \bm{r}_{i} -\bm{r}_{j}$. 
Hence, 
\begin{equation}
    \mathrm{W}_{i,\mu\nu} \propto \sum_j \frac{1}{|\bm{r}_{ji}|} \, r_{ji,\mu} \,\frac{\partial v(|\bm{r}_{ij}|)}{\partial |\bm{r}_{ji}|} \, r_{ji,\nu}
\end{equation}
is invariant under exchange of $\mu$ and $\nu$.
In the case of ML potentials with three-body descriptors, instead, we cannot assume that the atomic virials are symmetric, as already observed for general many-body potential \cite{Fan2015,Surblys2019,Boone2019,Surblys2021} }

\section{Results}
\subsection{Validation of MD-based methods}

\begin{figure}[t!]
    \centering
\includegraphics[width=\columnwidth]{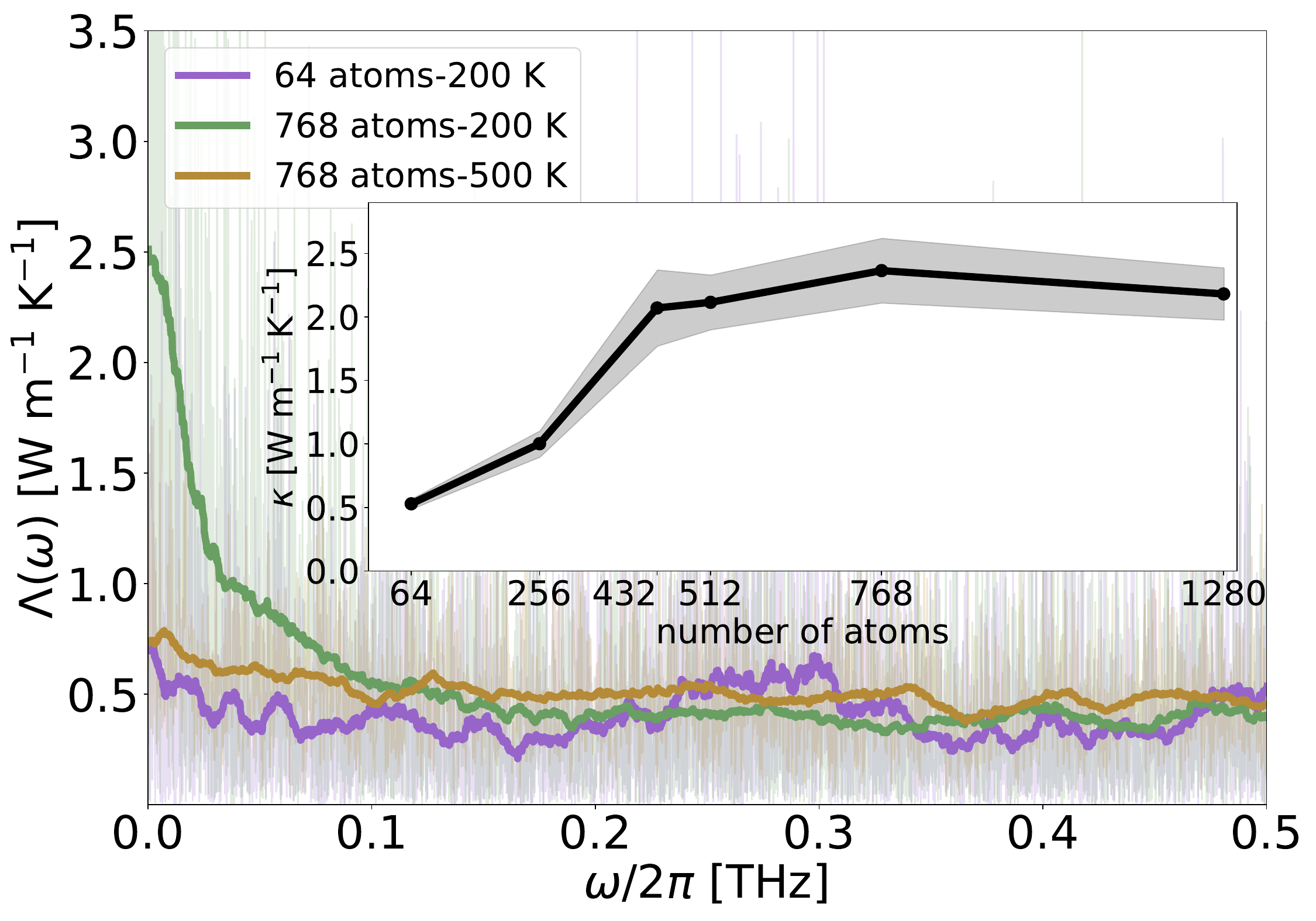}
    \caption{Low-frequency portion of the heat-flux reduced PSD for the $\gamma$ phase with 64-atom cell at $T=200$~K (violet), 768-atom cell at $T=200$~K (green) and 768-atom cell at $500$~K (gold), filtered with a window of 0.025 THz. The curve for the system of 768 atoms at $T=200$~K presents a steep peak (width $\approx 0.05$ THz) at very low frequencies ($<0.1$~THz).}
    \label{fig:lowfreqPSD_size}
\end{figure}

The application of the multivariate Green-Kubo method via MD (GKMD) is necessary for phases and temperatures where the Li-ion diffusion prevents any expansion into normal modes due to the lack of well-defined equilibrium positions of the atoms. However, at temperatures that are sufficiently low to inhibit Li-ion diffusion, one might wonder whether lattice methods (LM), like the Boltzmann transport equation (BTE), would be more appropriate. LM are indeed more straightforward from a computational perspective: once the relevant interatomic force constants are computed, $\kappa(T)$ is available at any temperatures at very low computational cost; furthermore, LM do not require the long dynamics necessary to sufficiently sample the slow modes through GKMD. 
An example of the latter behavior of GKMD is presented in Fig.~\ref{fig:lowfreqPSD_size}, which shows the heat-flux reduced PSD obtained via PBEsol-MLP on the $\gamma$ phase at different temperatures and cell sizes.
Resolving the narrow, steep peak at very low frequencies ($<0.1$~THz) -- that appears at $T=200$~K and is absent at larger $T$ -- requires MD simulations of several nanoseconds and determines the larger uncertainty on $\kappa$ for the $\gamma$ phase at low temperature. Notice that such a low-frequency peak can only be captured using a large simulation cell, due to the long wavelength of the modes involved, as shown by comparison with the 64-atom system (a size-dependent estimate of $\kappa$ and its uncertainty via cepstral analysis is shown in the inset to highlight finite-size effects).
Lattice methods, instead, allow to interpolate the modes in the Brillouin zone, hence sampling at once both high and very low frequencies, provided a large enough supercell is used to converge interatomic force constants.
Additionally, LM allow the inclusion of nuclear quantum effects (NQE) -- that, in the presence of a light element such as lithium, cannot be \textit{a priori} excluded \cite{MicheleNQE}, even at temperatures slightly above the Debye temperature, $T_D \approx 140~$K (see the SM~\cite{suppmat} for details on the computations) -- by simply considering quantum (Bose-Einstein) and classical (equipartition) occupations of the modes. However, the crucial drawback of LM is that, whenever the material is largely anharmonic, higher and higher orders of the normal-mode expansion should be computed and included in the calculation of $\kappa$. {In addition, LM are technically applicable only when it is possible to define a minimum of the potential energy surface, which is the case only for the $\gamma$ phase, but not for the $\beta$ and $\alpha$ phases, and in general for any other SSEs that have equivalent unoccupied Wyckoff sites. The latter phases would in fact require several expensive LM calculations on very large systems -- even in the absence of atomic diffusion -- to correctly sample the effects of disorder in the Li-ion sub-lattice and obtain reliable values for $\kappa$.} %

\begin{figure*}[t!]
    \centering
    \includegraphics[width=\linewidth]{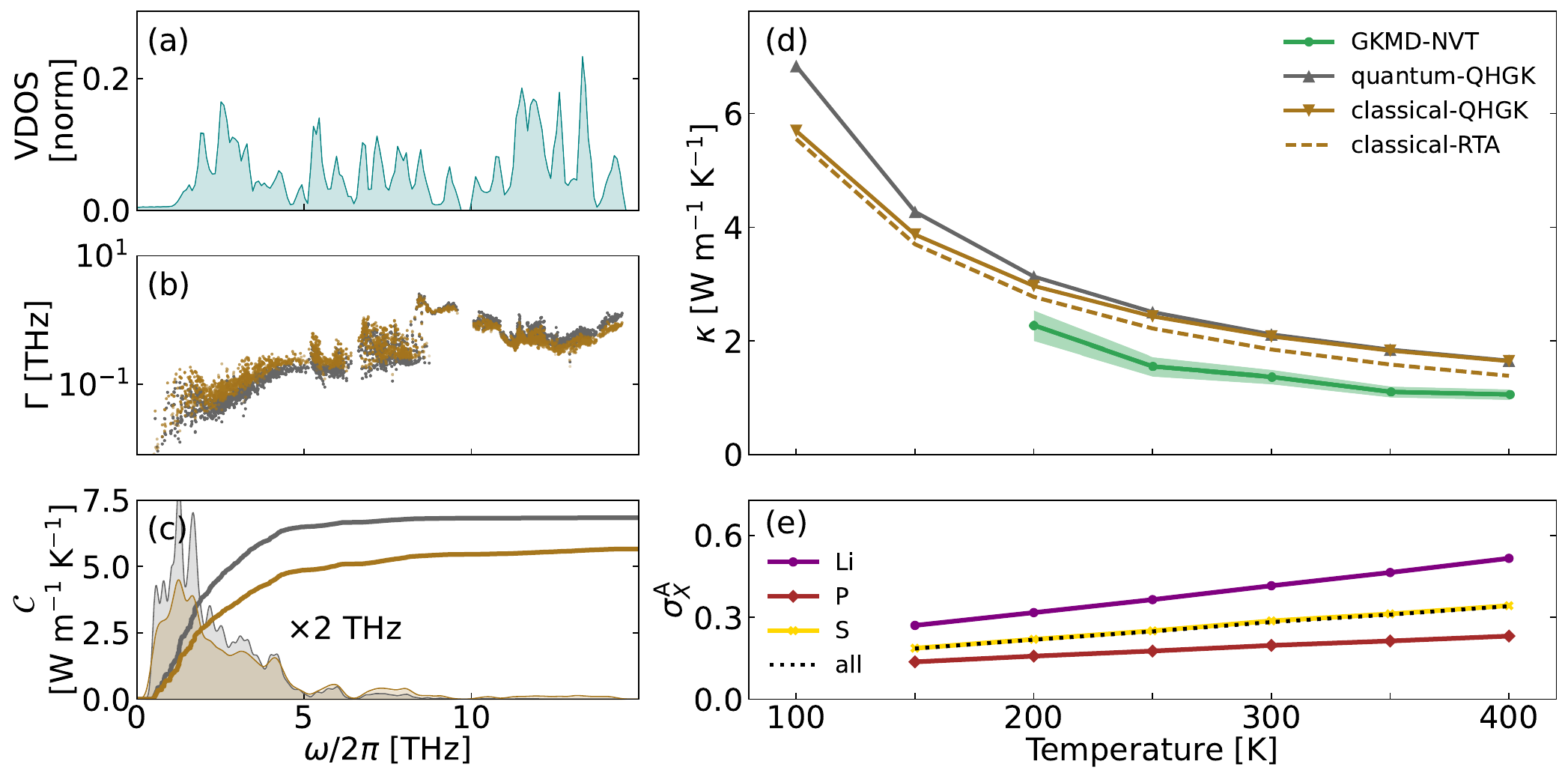}
    \caption{Lattice-method approach to thermal conductivity of the $\gamma$ phase at low $T$ for the PBE0 model. Vibrational density of states (a), phonon linewidths (b), and cumulative integral of lattice thermal conductivity, with its density (c) for {$T=100$~K}, for quantum (grey) and classical (brown) mode occupations. (d) $\kappa(T)$ via LM (on a $3\times 3\times 3$ supercell of the 16-atom primitive cell and $7\times 7\times 7$ interpolation) compared to the results from multicomponent GK theory for NVT EMD simulations. (e) Total and species-resolved degree of anharmonicity, Eq.~\eqref{eq:deg_of_anharm}, as function of $T$.}
    \label{fig:QHGK}
\end{figure*}

In the following, we show that, due to the strong anharmonic effects of \LiPS, and since NQE are shown to be largely negligible, GKMD has to be preferred with respect to LM even for the $\gamma$ phase, at $T \gtrsim 200$~K. 
To show this, we compute $\kappa$ adopting {LM with the recent formulations described in Refs.~\onlinecite{Simoncelli2019, Isaeva2019}, which, besides recovering the BTE, are also able to account for effective corrections due to \textit{interband} transitions between vibrational modes. To do so, we adopt the quasi-harmonic Green-Kubo (QHGK) formalism~\cite{Isaeva2019,NoteFiorentino}, as implemented in $\kappa$ALD$o$ \cite{Barbalinardo2020}.}

Figure \ref{fig:QHGK} shows LM results for PBE0 model, which displays the best overall accuracy against experiments for several physical observables~\cite{Gigli2023}. The vibrational density of states (VDOS), Fig.~\ref{fig:QHGK}(a), and the mode bandwidths $\Gamma$, plotted at $T=200$ K in Fig.~\ref{fig:QHGK}(b), are computed from finite differences using the \textsc{librascal}-ASE calculator \cite{librascal,ASE2017}, interfaced with $\kappa$ALD$o$. 
The per-mode contribution to the thermal conductivity, and its cumulative value, $\mathcal{C}(\omega)$, are shown in Fig.~\ref{fig:QHGK}(c) for $T=200$~K. We stress that $\mathcal{C}(\omega)$ stems from a modal representation of heat transport. Therefore it is must \textit{not} be misinterpreted as an energy flux PSD: in fact, the lattice thermal conductivity can be either estimated in terms of the $\omega\to 0$ limit of the energy-flux PSD, or as the $\omega\to\infty$ limit of the cumulant function $\mathcal{C}(\omega)$.
See SM~\cite{suppmat} for further tests and computational details along with a study of the temperature dependence of the spectral decomposition of $\kappa$.

The results for $\kappa(T)$ for quantum and classical mode occupations are shown and compared with GKMD in Fig.~\ref{fig:QHGK}(d).  The increase in $\kappa$ due to interband coupling \cite{Isaeva2019,Simoncelli2019} is also highlighted by direct comparison of the QHGK conductivity $\kappa_\mathrm{QHGK}$ with the single-mode relaxation-time approximation $\kappa_\mathrm{RTA}$---which accounts for \textit{intra}band relaxation only. Even the correction to RTA introduced by the inversion of the full scattering matrix in the BTE cannot be responsible for the difference between LM and MD results, since it amounts to less than 5\% of the $\kappa_\mathrm{RTA}$, as detailed in SM~\cite{suppmat}, Sec.~VIII.

As it is often the case \cite{Pegolo2022,Puligheddu2019}, the thermal conductivity obtained via lattice methods is larger than that computed by GKMD, even when the classical distribution is employed, likely because of higher-order contributions to normal-mode scattering. 
To quantify the role of anharmonicity, since computing higher-order phonon scattering rate on the same supercell would be intractable, we employ the \textit{degree of anharmonicity}, 
\begin{equation}
    \sigma^\mathrm{A}_{X} \equiv \sqrt{\frac{\sum_{i\in X}\sum_{\mu} \langle|F_{i,\mu}(\bm{R}) - F^{(2)}_{i,\mu}(\bm{R})|^2\rangle}{\sum_{i\in X}\sum_{\mu} \langle|F_{i,\mu}(\bm{R})|^2\rangle}} \label{eq:deg_of_anharm}
\end{equation}
introduced by Knoop \textit{et al.}~\cite{Knoop2020}. Here, $F_{i,\mu}(\bm{R})$ is the force acting on atom $i$ along Cartesian component $\mu$, a function of all atomic coordinates $\bm{R}\equiv (\bm{r}_1,\ldots,\bm{r}_N)$, and $F^{(2)}_{i,\mu}(\bm{R}) = - \sum_{j,\beta} \Phi_{i,\mu}^{j,\nu} \Delta r_{j,\nu}$ is its harmonic approximation, $\Phi_{i,\mu}^{j,\nu}$ being the set of force constants and $\Delta \bm{r}_{j}$ the deviation from equilibrium of atom $j$. $\sigma^\mathrm{A}_X$ is reported in Fig.~\ref{fig:QHGK}(e) for different subsets of atoms $X=\mathrm{Li,P,S}$. The values are those typical of highly anharmonic materials like rocksalts, perovskites or chalcopyrites \cite{Knoop2023}. For the sake of comparison, at room temperature, silicon, a largely harmonic material, has significantly smaller $\sigma^\mathrm{A}_\mathrm{Si} \approx 0.15$ \cite{Knoop2020}. Both Li and S atomic vibrations are very anharmonic. %
The reason is quite intuitive for Li, as Li ions are prone to diffuse at larger $T$, while for S it is a consequence of the S librations that are a characteristic feature of \LiPS. %
{Overall, this analysis suggests that LM predict values of $\kappa$ that are of the same order of magnitude than GKMD, yet consistently higher. Incorporating higher-order contributions to normal mode scattering, which generally reduce $\kappa$ \cite{Pegolo2022}, might align LM predictions more closely with GKMD results.}

\subsection{Green-Kubo molecular dynamics}

Having extensively justified, even at low $T$, the need for the GKMD method to calculate $\kappa$, we now discuss the results obtained for the thermal conductivity of the $\gamma$, $\beta$ and $\alpha$ phases of \LiPS~over a wide range of temperatures, for MLPs targeting PBEsol, r$^2$SCAN and PBE0 functionals. EMD simulations on 768-atom cells are run with i-PI \cite{Kapil2019} in the NpT ensemble, with stochastic velocity rescaling (SVR) thermostat \cite{buss+07jcp} and a generalized Langevin barostat \cite{ceri+09prl} on the cell volume to target zero pressure. {In contrast with other types of thermostats, SVR is able to reproduce NVE results for dynamical properties featuring characteristic time scales of several picoseconds, such as the shear viscosity of water or the dynamic structure factor of a polymer chain in explicit water \cite{Basconi2013}. Furthermore, the SVR thermostat was recently shown to recover the results of GKMD thermal conductivity of simple superionic materials from NVE simulations, even at temperatures low enough to suppress ionic diffusion (see, e.g., the case of CaF$_2$ in Fig.~S10 of the SM of Ref.~\onlinecite{Grasselli2022}).} We have {also} verified that the barostat, with the chosen parameters, does not affect the statistical properties of correlation functions and GK estimates {from NpT simulations, when compared to NVT} (see the SM \cite{suppmat}, Fig.~S8). EMD trajectories are at least 2-ns long, and energy and mass fluxes are sampled every 10 fs, that is sufficient to avoid aliasing in the PSD.
\begin{figure}
    \centering
    \includegraphics[width=\columnwidth]{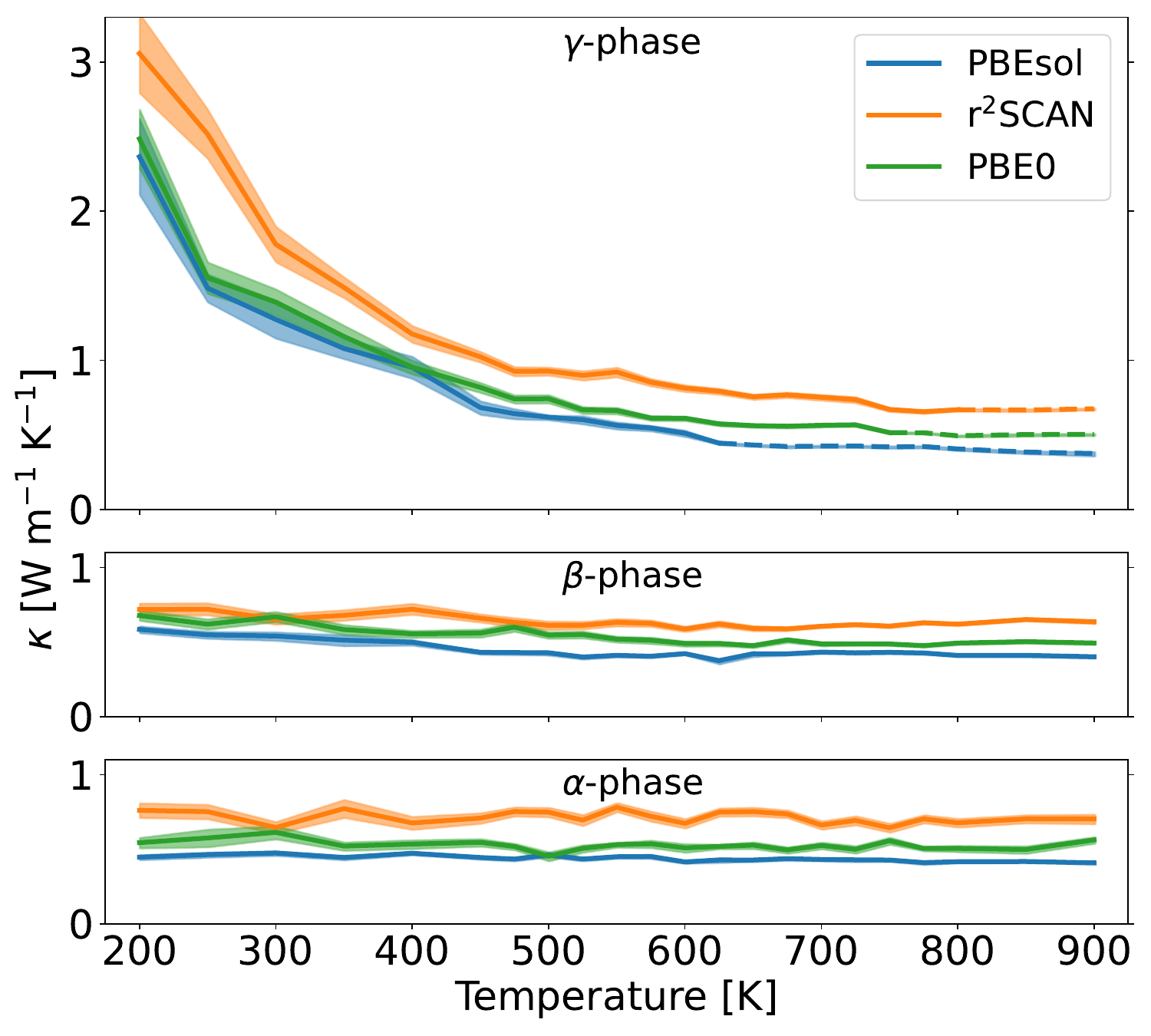}
    \caption{Temperature behavior of the thermal conductivity of \LiPS, $\kappa(T)$, for the $\gamma$, $\beta$ and $\alpha$ phases and all the ML-models, from multicomponent Green-Kubo theory. Estimates (lines) and uncertainties (shaded areas) are obtained via multivariate cepstral analysis. The dashed lines in the top panel represent the regime where simulations, despite being initialized in the $\gamma$ phase, undergo a transition to a mixed $\alpha/\beta$ phase, characterized by different ordering of \ce{PS4} tetrahedra~\cite{Gigli2023}. Notice the different vertical scale of top panel.}
    \label{fig:kappa_phases_functionals}
\end{figure}
Imperfect ergodicity and the constraint on cell shape make simulations dependent on the phase in which the system is initialized. In Ref.~\onlinecite{Gigli2023}, we have shown that simulations initialized in the nonconductive $\gamma$ phase undergo a structural phase transition to a mixed $\alpha/\beta$ phase that is Li-ion conductive, whenever simulations are run beyond a critical temperature, $T_c$, that strongly depends on the DFT functional ($T_c\approx$ 625 K for PBEsol, 750 K for PBE0, and 800 K for r$^2$SCAN).
Experimentally, the $\gamma$-to-$\beta$ transition (structural transition from symmetry group $Pmn2_1$ to $Pnma$) occurs at $573~$K, while the transition to the $\alpha$-phase ($Cmcm$ \cite{kaup_impact_2020}) occurs at $746~$K \cite{HOMMA201153}.

Figure~\ref{fig:kappa_phases_functionals} shows the results of the multicomponent cepstral analysis.
Following previous studies of heat transport in ionic conductors \cite{Agne2022, Bernges2022}, we introduce two distinct heat-transport carriers, i.e.~propagating and diffusing normal modes: propagating modes (also known as \textit{propagons}), possess a nonvanishing group velocity and transport heat on a length scale usually larger than the interatomic distance via \textit{intraband} scattering following conservation of total momentum. In contrast, diffusing modes (also known as \textit{diffusons}) do not propagate and transport heat much more locally thanks to the \textit{interband} coupling between different vibrational modes that are almost degenerate in energy and possess a finite lifetime.
The simulations initialized in the $\gamma$ phase (top panel) show a low-temperature crystal-like behavior, whereby $\kappa$ decays with $T$, which is typical of heat transport mediated by phonon propagation.
In contrast, simulations initialized in the $\beta$ and $\alpha$ phases (mid and bottom panels, respectively) show an effective ``glass-like'' behavior \cite{Cheng2021} whereby $\kappa$ is in practice independent of $T$: even at very low temperature, the disorder stemming from the large availability of empty Wyckoff sites for Li-ions in these phases (which promotes Li-diffusion at larger $T$) hinders phonon propagation in favor of diffusons, within the so-called Allen-Feldman regime \cite{Allen1993}, making these phases act like glasses, despite the overall crystalline arrangement of the \ce{PS4} tetrahedra.
Notice that the value of $\kappa$ obtained from simulations initialized in the $\gamma$ phase but equilibrated at $T>T_c$ matches the value obtained in simulations initialized in the $\alpha$ and $\beta$ phases. The regime $T>T_c$ is represented by the dashed line in the top panel of \cref{fig:kappa_phases_functionals}.

For each $T$ and phase, the largest values of $\kappa$ are observed for r$^2$SCAN model, followed by PBE0, and finally by PBEsol. We have verified that the mere difference in the equilibrium volumes predicted by the different functionals cannot explain \textit{per se} the significant difference between the estimates for $\kappa$ (see the SM \cite{suppmat}, Sec.~VII).
Interestingly, the largest \textit{electrical} conductivities are instead found for PBEsol \cite{Gigli2023}: Li-ion diffusion affects the structure of the SSE by favoring disorder and inducing diffuson-like behavior typical of glass-like systems, even when the solid matrix (here composed by \ce{PS4} tetrahedra) maintains crystalline order. 
The lower bulk modulus predicted for the $\gamma$ phase by PBEsol (see the SM \cite{suppmat}, Sec.~II) also indicates softer bonding, usually promoting diffuson-like thermal transport \cite{Hanus2019}. Nevertheless, it would be improper, if not wrong, to speak of a quantitative relation between ionic and thermal conduction, especially on a local scale \cite{Agne2022}. For instance, recent studies on argyrodites, another class of SSEs, highlight no significant correlation between the magnitude of ionic and thermal conductivity \cite{Bernges2022}.

\section{Conclusions}

{Our thorough analysis of the nondiffusive, low-$T$ $\gamma$ phase shows that one must be very careful when using both lattice methods and molecular dynamics to study the thermal properties of SSEs: On one hand, lattice methods are not suited for systems with degrees of disorder, like those with empty Wyckoff sites where several equivalent minima of the potential energy surface are allowed. Moreover, the presence of large anharmonicity in \LiPS~even at low $T$ might make calculations up to three-phonon scattering not entirely reliable. On the other hand, MD-based methods need very long simulations to correctly sample the scattering of long-living, low-frequency phonons.} Furthermore, large-scale simulations are needed, as finite-size effects tend to lower the estimate of $\kappa$ with respect to converged values \cite{Grasselli2022}, making ML potentials essential for accurate simulations.

We also stress that MD-based calculations are in general tolerant to small inaccuracies of the force field (as long as the dynamics is correctly reproduced on average \cite{Pastore1991}), in contrast to lattice methods, where very accurate interatomic force constants are required. This feature further leans towards the use of GKMD, whenever ML potentials, usually trained on lower-order energy derivatives, are used and may not be able to reproduce high-order force constants, as shown for glasses in Ref.~\cite{Pegolofrontiers2024}.
The use of GKMD to study these systems is also justified by the observation that nuclear quantum effects are negligible in the typical operating temperature regime (beyond 200 K) of SSEs.
{The use of MD also permits the study of other quantities crucial for the characterization of SSEs, such as the electrical conductivity, as well as conduction mechanisms. For instance, in Ref.~\cite{Gigli2023}, it has been shown that the ML-PBE0 model performs better than ML models targeting other \textit{ab initio} functionals in predicting the charge-transport properties of \LiPS.}

Finally, in the Li-diffusive $\alpha$ and $\beta$ phases of \LiPS, the thermal conductivity is characterized by a glass-like behavior, with no significant dependence on the temperature nor on the specific phase, and values in line with recent experimental measures on amorphous lithium thiophosphates, i.e.~in the range 0.45–0.70 W m$^{-1}$ K$^{-1}$  \cite{Cheng2021}: the crystalline arrangement of \ce{PS4} groups is not sufficient to preserve the propagonic crystal-like behavior of $\kappa(T)$, which is hindered by the Li-ion disorder, induced by Li-ion diffusion and, even at low temperature, by the large availability of empty Wyckoff sites for Li ions, which are randomly arranged. {Similarly, low values  and a glass-like behavior of $\kappa$, indicating a prevalence of diffuson-like thermal transport, have also been found in experimental works on similar materials, such as Na$_3$PS$_4$ \cite{Bernges2022Na3PS4}. }

In summary, our study demonstrates that the combination of MLPs with GKMD is a promising approach for accurately calculating thermal conductivity at an \textit{ab initio} level in lithium-based electrolytes. Large anharmonicity and ion diffusion make, in general, GKMD results more reliable than lattice methods for SSEs, once the multi-component analysis is correctly applied. This may be relevant for the prospective application to SSEs of high-throughput screening for thermal conductivity, that recently attracted great interest thanks to new semiautomated frameworks \cite{Knoop2023PRB,Knoop2023,Meiirbek2023}, in analogy to what has been pursued for the electrical conductivity \cite{Sendek2017,Muy2019,he2020high,Kahle2020} of SSEs.
In particular, the possibility of computing the energy-flux time series in post processing also allows one to access thermal conductivity from the MD trajectories already produced to investigate the electrical conductivity which may be beneficial to jointly screen for both heat and charge transport of SSEs.

\section*{Data availability}

All the relevant data, scripts, and input files that support the results of this work are available on the Materials Cloud platform \cite{Talirz2020} at [\href{https://doi.org/10.24435/materialscloud:nv-1g}{https://doi.org/10.24435/materialscloud:nv-1g}]. The heat-flux time series employed in the multicomponent Green-Kubo analysis and the interatomic force constants used in the Boltzmann transport equation are available on Zenodo \cite{zenodo} at [\href{https://zenodo.org/records/10977594}{https://zenodo.org/records/10977594}]
\section*{Acknowledgments}

We thank A.~Goscinski, P.~Pegolo and H.~C.~T\"{u}rk for technical help and a critical review of an early version of this manuscript. We thank E.~Drigo and F.~Knoop for insightful discussions on the quantification of anharmonicity in solids.

D.T., L.G.~and M.C. acknowledge funding from the Swiss National Science Foundation (SNSF) under the Sinergia project CRSII5\_202296 and support from the MARVEL National Centre of Competence in Research (NCCR) for computational resources. F.G.~and M.C.~acknowledge funding from the European Research Council (ERC) under the European Union’s Horizon 2020 research and innovation programme Grant No.~101001890-FIAMMA. F.G.~also acknowledges funding from the European Union's Horizon 2020 research and innovation programme under the Marie Sk\l{}odowska-Curie Action IF-EF-ST, grant agreement No.~101018557-TRANQUIL. This work was supported by grants from the Swiss National Supercomputing Centre (CSCS) under the projects s1092 and s1219 and by the Platform for Advanced Scientific Computing (PASC) project ``Machine learning for materials and molecules: toward the exascale''.

\end{document}